# Reply to "Non-relativistic proofs of the spin-statistics connection," by Shaji and Sudarshan.

**Murray Peshkin**
Physics Division, Argonne National Laboratory, Argonne, IL 60439

E-mail:  **peshkin@anl.gov**

## ABSTRACT

I earlier proved under stated assumptions of nonrelativistic quantum mechanics, that identical spin-zero particles with no internal degrees of freedom must be bosons. Shaji and Sudarshan have criticized that proof, asserting that it is "based on single-valuedness under rotation of the wave functions of systems of identical particles." No such assumption was used. Here, to remove all doubt, I exclude the fermion possibility through a proof in which even the existence of rotation is not used. Following that, I assume that the Hilbert space is invariant under rotation to exclude other symmetries which correspond neither to bosons nor to fermions. In that proof no issue of single-valuedness can arise.

I. INTRODUCTION

Shaji and Sudarshan [1] have asserted that my proof [2] (hereafter called MP1) that spin-zero particles must be bosons was "based on the single-valuedness under rotation of the wave functions of systems of identical spin-zero particles." In making that assertion, they do not specify where in MP1 the assumption of single-valuedness was used. In fact it was not used. Here, to remove all doubt, I first give a new proof, one that does not assume even the existence of rotation, that two identical spin-zero particles with no degrees of freedom other than the spatial ones cannot be fermions. Following that, I consider the possibility of a mixed symmetry in which the particles are neither fermions nor bosons. That possibility is excluded by a simple argument that does rely upon the Hilbert spaced being invariant under rotation, but it does not involve any specifics of rotation operators and no question of single-valuedness pertains.

The exclusion of the mixed symmetries was proved in MP1 in a very different way, by carrying out a rotation that in effect exchanges a pair of particles. That proof explicitly allowed for the possibility of multiple-valuedness under rotation by introducing an undetermined phase in its Eq.(A4). Section V below includes an outline of the proof in MP1 that shows how the undetermined phase enters the reasoning.

This paper is self contained and independent of MP1, but it is not complete. The extension to many particles and the details of the use of rotation are not repeated here.

II. ASSUMPTIONS

The principal idea of this approach, and the only non-standard assumption of nonrelativistic quantum mechanics used here, is the specialization to spin zero of an assumption introduced by Leinaas and Myrheim [3] and developed by Berry and Robbins [4,5]. For two identical spin-zero particles it is assumed that the spatial variables in the wave functions should be the unordered pairs of points $\{\mathbf{r}_1, \mathbf{r}_2\}$, for which $\{\mathbf{r}_2, \mathbf{r}_1\}$ is the same point in the configuration space as is $\{\mathbf{r}_1, \mathbf{r}_2\}$. More generally, for N identical spin-zero particles, the wave function should depend upon the unordered multiplet $\{\mathbf{r}_1, \mathbf{r}_2, ... \mathbf{r}_N\}$. This assumption was motivated by the notion that the coordinates in quantum mechanics should represent the spatial observables. It stands in contrast to the usual practice of writing many-body wave functions as functions of the coordinates of labeled particles in defiance of the non-existence of identifying labels on identical particles, and then symmetrizing or antisymmetrizing the wave functions for reasons not justified within nonrelativistic quantum mechanics.

I make the usual assumption that the wave function for spin-zero particles with no internal variables has a single component which depends only upon spatial variables, and that it is a single-valued function of those spatial variables. In MP1, I further assumed that the wave function is scalar under rotation. That additional assumption is not used here. Single-valuedness as a function of the spatial variables is unavoidable in the conventional case if the domain of those variables is simply connected, *i.e.* if there are no closed loops that cannot be shrunk to a point. The same is true of the unordered pairs used here. In view of the question that has been raised about single-valuedness under rotation, it is perhaps useful to recall that the Dirac wave function, like all wave functions, is a single-valued function of its spatial variables although multiple-valued under rotation.

I also make essential use of the assumption that wave functions must be continuous under infinitesimal displacements of the particle coordinates because of the second derivatives in the Hamiltonian. That assumption is really part of conventional nonrelativistic quantum mechanics because a discontinuous wave function necessarily has infinite energy $E = \langle \psi | H | \psi \rangle$, but it is not usually emphasized.

The theory presented here and in MP1 is not new. It is equivalent to the standard theory in which one imposes symmetry of the wave function for reasons outside of nonrelativistic quantum mechanics. The present approach is motivated by the desire to have a simple and transparent proof within nonrelativistic quantum mechanics which relies only upon stated assumptions that seem acceptable and give some simple insight into what underlies the spin-statistics connection.

## III. THE UNORDERED PAIRS

The domain of the unordered pairs is half of the full six-dimensional space that would apply to distinguishable particles, for which $\mathbf{r}_1$ and $\mathbf{r}_2$ are separately observable. For distinguishable particles, one has center-of-mass and relative coordinates $\mathbf{R} = (\mathbf{r}_1 + \mathbf{r}_2)/2$ and $\mathbf{r} = \mathbf{r}_2 - \mathbf{r}_1$, the domain of each being the full three-dimensional space. For identical particles, the relative coordinate of the unordered pair is an undirected line segment joining $\mathbf{r}_1$ and $\mathbf{r}_2$. It has a magnitude and an axis in space, but no direction along that axis. The polar coordinates of such an axis, taken to be to be one-to-one with the axes and therefore with the unordered pairs, have the domain

$$0 \le \vartheta \le \frac{\pi}{2}$$
$$0 \le \varphi < 2\pi \quad \text{for} \quad \vartheta < \frac{\pi}{2} \tag{1}$$
$$0 \le \varphi < \pi \quad \text{for} \quad \vartheta = \frac{\pi}{2}$$

where the choice of the polar axis, here the *z* axis, is of course arbitrary. Similarly, one could add an arbitrary constant to the bounds of $\varphi$.

Equivalently, one can define the relative coordinate $\mathbf{r}$ by

$$\mathbf{r} = \begin{cases} \mathbf{r}_2 - \mathbf{r}_1 & \text{where} \begin{cases} z_2 > z_1, & \text{or} \\ z_2 = z_1 \text{ and } y_2 > y_1, & \text{or} \\ z_2 = z_1 \text{ and } y_2 = y_1 \text{ and } x_2 \geq x_1 \end{cases} \\ \mathbf{r}_1 - \mathbf{r}_2 & \text{elsewhere} \end{cases} \quad (2)$$

The domain $D$ of the so-defined relative $\mathbf{r}$ is the half-space consisting of points with positive $z$ and half of the $z=0$ plane.

$$D = [z > 0] \oplus [z = 0 \text{ and } y > 0] \oplus [y = z = 0 \text{ and } x \geq 0] \quad (3)$$

When a non-null vector $\mathbf{r}$ lies within the domain $D$, $-\mathbf{r}$ does not. Eq.(2) is uniquely invertible to give

$$\{\mathbf{r}_1, \mathbf{r}_2\} = \left\{ \left(\mathbf{R} + \frac{\mathbf{r}}{2}\right), \left(\mathbf{R} - \frac{\mathbf{r}}{2}\right) \right\} \quad (4)$$

The mapping in Eq.(2) of the unordered pairs onto $(\mathbf{R},\mathbf{r})$ is discontinuous at the $z_1 = z_2$ plane, and only there. Consider two neighboring pairs $\{\mathbf{r}_1,\mathbf{r}_2\}$ and $\{\mathbf{r}_1',\mathbf{r}_2'\}$, defined by

$$\{\mathbf{r}_1, \mathbf{r}_2\} = \{(x, y, z + \varepsilon), (-x, -y, z)\} \Leftrightarrow \mathbf{r} = (2x, 2y, \varepsilon)$$
$$\{\mathbf{r}_1', \mathbf{r}_2'\} = \{(x, y, z - \varepsilon), (-x, -y, z)\} \Leftrightarrow \mathbf{r}' = (-2x, -2y, \varepsilon) \quad (5)$$

with positive $\varepsilon$. In the limit $\varepsilon \to 0$, the relative coordinate has a jump discontinuity except at $x=y=0$. That there should be some discontinuity is unavoidable since Eq.(2) implies an ordering of vectors in three dimensions that could be used to map three dimensions onto one. The physical consequences of the discontinuity are addressed in Section IV below.

The representation of the unordered pairs by vectors in the half-space used here is that of Leinaas and Myrheim, except that I make the domain of the relative coordinate simply connected and one-to-one with the unordered pairs, whereas Refs.(3-5) achieve one-to-one representation by identifying pairs of points in a-multiply-connected domain of $\mathbf{r}$.

## IV. EXCHANGE AND SYMMETRY

In conventional practice, where the particles are treated as if they were distinguishable, a general wave function can be written as

$$\Psi(\mathbf{r}_1, \mathbf{r}_2) = \psi(\mathbf{R}, r, \vartheta, \varphi) = \sum_{\ell, m, n} a_{\ell m n}(\mathbf{R}) g_n(r) Y_{\ell m}(\vartheta, \varphi), \quad (6)$$

where the $g_n(r)$ are some complete orthonormal set and $(\vartheta,\varphi)$ are the angles of the relative **r**. The domain of **r** is the entire three-dimensional space and the $Y_{\ell m}(\vartheta,\varphi)$ for all integer values of $\ell$ are a complete orthonormal set. In that case, the exchange $\mathbf{r}_1 \leftrightarrow \mathbf{r}_2$ is equivalent to a rotation of the relative **r** about a perpendicular axis to carry **r** into $-\mathbf{r}$, and one can prove that restricting the sum in Eq.(6) to only even or only odd values of $\ell$ is equivalent to assuming that

$$\Psi(\mathbf{r}_2,\mathbf{r}_1) = (-1)^\ell \Psi(\mathbf{r}_1,\mathbf{r}_2). \tag{7}$$

Bosonic behavior, with wave functions having even $\ell$ in their relative motion, is associated with symmetry of the wave function under exchange. Fermionic behavior is associated with odd $\ell$ and antisymmetry of the wave function under exchange.

No such proof exists in the geometry used here for identical spin-zero particles. The exchange $\mathbf{r}_1 \leftrightarrow \mathbf{r}_2$ has no relation to any rotation of the relative **r**, whose domain does not include $-\mathbf{r}$. The wave function is always symmetric under $\mathbf{r}_1 \leftrightarrow \mathbf{r}_2$ because it depends only upon the unordered pair, or equivalently upon the relative **r** as defined above on the half-space. The $Y_{\ell m}(\vartheta,\varphi)$ for even and for odd $\ell$ are separately complete on the half space. Under the assumptions used to this point, spin-zero particles may be either bosons or fermions, according to the choice of even or odd $\ell$. For example, the wave function

$$\psi(\mathbf{R},\mathbf{r}) = (x+iy)e^{-r^2}e^{-R^2} \tag{8}$$

is symmetric under $\mathbf{r}_1 \leftrightarrow \mathbf{r}_2$, as are all functions of **R** and **r**, but it has $\ell=1$.

IV. CONTINUITY OF THE WAVE FUNCTIONS AND
THE ELIMINATION OF SPIN-ZERO FERMIONS.

The choice of even or of odd $\ell$ defines a Hilbert space for the two-particle wave functions. The two lead to different physics. Here I will show that the choice of odd $\ell$ is excluded by the continuity condition.

As discussed in Section II above, wave functions must be continuous under infinitesimal displacements of $\mathbf{r}_1$ and $\mathbf{r}_2$. Then Eq.(5) requires that in the limit $\varepsilon \to 0$ from above,

$$\psi\left(\mathbf{R},r,\frac{\pi}{2}-\varepsilon,\varphi \pm \pi\right) \to \psi\left(\mathbf{R},r,\frac{\pi}{2}-\varepsilon,\varphi\right). \tag{9}$$

$$\sum_{\ell,m,n} a_{\ell m n}(\mathbf{R})g_n(r)(-1)^m Y_{\ell m}(\frac{\pi}{2},\varphi) = \sum_{\ell,m,n} a_{\ell m n}(\mathbf{R})g_n(r)Y_{\ell m}(\frac{\pi}{2},\varphi) \tag{10}$$

$$a_{\ell mn}(\mathbf{R})Y_{\ell m}(\frac{\pi}{2},\varphi) = 0 \text{ for all odd } m. \tag{11}$$

However, $Y_{\ell m}(\frac{\pi}{2},\varphi) = 0$ if and only if $(\ell\text{-}m)$ is odd. Then

$$a_{\ell mn}(\mathbf{R}) = 0 \text{ for all } \mathbf{R} \text{ and all } n, \text{ when } \ell \text{ and } m \text{ are both odd.} \tag{12}$$

All $m$ are allowed for even $\ell$.

Therefore two spin-zero particles are allowed to be bosons with even relative $\ell$, but not to be fermions with odd relative $\ell$. The $Y_{\ell m}$ with even $\ell$ are an acceptable basis set to describe the physics. A Hilbert space built on the $Y_{\ell m}$ with odd $\ell$ would have to be incomplete, containing only even values of $m$. For example, all continuous wave functions in such a space vanish when the relative $z=0$. In other language, the Hamiltonian is not self-adjoint in such a Hilbert space. In proving all this, no assumption was made about even the conceptual existence of rotation, although rotations do of course exist.

## V. ROTATION AND THE ELIMINATION OF MIXED $\ell$

Other mathematical possibilities exist in addition to odd $\ell$ alone (fermions) and even $\ell$ alone (bosons). One could build a Hilbert space on a complete orthonormal set consisting of the $Y_{\ell m}$ with all the odd $\ell \geq |m|$ for some or all odd values of $m$ and the $Y_{\ell m}$ with even $\ell$ for all the remaining $m$. More generally, one could use any orthonormal set of linear combinations of the $Y_{\ell m}$ that avoids having a contribution from odd $\ell$ with odd $m$. Each such basis set defines a different theory with possibly different physical consequences. They are all physically unacceptable because their Hilbert spaces are not invariant under rotation. To be invariant under rotation, they must include all $m$ or none for a each $\ell$. Otherwise, the Hilbert space depends upon the choice of the $z$ axis.

The last two sentences in the previous paragraph contain the only appearance of rotation in the proof given here that spin-zero particles must be bosons. In MP1, I eliminated wave functions that mix $Y_{\ell m}$ with even with odd values of $\ell$ in an alternative way that illuminates the trouble with mixed symmetry somewhat differently. There I supposed that one can have basis states $|\mathbf{r}_0\rangle$ for all $\mathbf{r}_0$ in the Domain $D$ of Eq.(3). Carrying one such $|\mathbf{r}_0\rangle$ to within $\varepsilon$ of the $z=0$ plane by rotation around an axis perpendicular to $\mathbf{r}_0$ in two directions yields two states $|\mathbf{r_0}'\rangle$ and $|\mathbf{r_0}''\rangle$ that are within $\varepsilon$ of representing the same unordered pair, so that in the limit $\varepsilon \rightarrow 0$, one must have

$$|\mathbf{r_0}'\rangle = e^{i\delta}|\mathbf{r_0}''\rangle \qquad (13)$$

for some $\delta$. Properties of the rotation matrices were then used to show that Eq.(13) is violated if $|\mathbf{r_0}\rangle$ contains both even and odd $\ell$. The assumption that $\delta$ equals zero would have corresponded to single-valuedness of the wave function under rotation, but that assumption was not used.

## VI. SUMMARY

Conventional nonrelativistic quantum mechanics allows two identical spin-zero particles with no other degrees of freedom than their spatial ones to be bosons with only even $\ell$ in their relative motion, fermions with only odd relative $\ell$, or possibly something else with a mixture of even and odd relative $\ell$. It has been shown here without recourse to any assumption about rotation that the Leinaas-Myrheim assumption that the argument of the wave function is an unordered pair of points, coupled with the assumption of continuity, eliminates the fermion possibility. Adding the assumption that the Hilbert space should be invariant under rotation, which requires that it includes the $Y_{\ell m}$ for all $m$ or none for a given value of $\ell$, eliminates mixtures of even and odd relative $\ell$. Then the only remaining possibility is that the particles have only even $\ell$ and are bosons.

As was noted in MP1, the resulting theory is equivalent to the conventional theory wherein the individual particles are given labels and the configuration space is taken as the full six-dimensional space but the wave functions are restricted to force the particles to be bosons. Wave functions of the conventional theory are obtained from those used here by extending the domain of $\mathbf{r}$ to the entire space and making the wave functions even under inversion of $\mathbf{r}$ and renormalizing them by a factor $\sqrt{2}$. All physical operators are also even under inversion of $\mathbf{r}$. Therefore every matrix element of every observable is the same in the present theory as in the conventional theory and the two are physically equivalent to each other. In that sense, the assumption that wave functions should depend only upon the unordered pairs provides a justification for imposing symmetry of the wave function under exchange in the conventional theory.

It was also shown in MP1 that adding an assumption of asymptotic separability, *i.e.* the assumption that the relative motion of two particles is not influenced by other particles at infinite distance, enables generalization of this result to any number of spin-zero particles. The relative motion of each pair contains only even $\ell$. That proof will not be repeated here.

Finally, I note that Shaji and Sudarshan refer in their abstract to criticisms of MP1 by Allen and Mondragon [6]. Those criticisms, which relate to issues not connected with rotation, have been rebutted elsewhere [7].

## ACKNOWLEDGMENTS

I thank Fritz Coester for his relentless constructive criticisms. This work is supported by the U.S. Department of Energy, Nuclear Physics Division, under Contract No. W-31-109-ENG-38.